\documentstyle[12pt,preprint,psfig]{aastex}

\newcommand{\hmpc}{$h^{-1}$\,Mpc}
\newcommand{\be}{\begin{equation}}
\newcommand{\e}{\end{equation}}
\newcommand{\f}{\frac}

\shorttitle{Absorption systems: Comparison of simulations
with observations}

\begin{document}

\title
{Semi analytic approach to understanding the distribution 
of neutral hydrogen in the universe: Comparison of simulations
with observations}
\author{T. Roy Choudhury \\ tirth@iucaa.ernet.in} 
\author{R. Srianand \\ anand@iucaa.ernet.in} 
\and 
\author{T. Padmanabhan \\ paddy@iucaa.ernet.in}
\affil{IUCAA, Post Bag 4, Ganeshkhind, Pune 411 007, India.}


\begin{abstract}
Following Bi \& Davidsen (1997), we perform one dimensional semi analytic 
simulations along the lines of sight to model the intergalactic medium (IGM). 
Since this procedure is computationally 
efficient in probing the parameter space -- and reasonably accurate -- we 
use it to recover the values of various parameters related to the IGM 
(for a fixed background cosmology) 
by comparing the model predictions with different observations.
For the currently favoured LCDM model  
($\Omega_m=0.4$, $\Omega_{\Lambda}=0.6$ and $h=0.65$), we obtain, 
using
statistics obtained from the transmitted flux, 
constraints on (i) the combination $f=(\Omega_B h^2)^2/J_{-12}$, where 
$\Omega_B$ is the baryonic density parameter and $J_{-12}$ is the 
total photoionisation rate in units of $10^{-12}$s$^{-1}$, 
(ii) 
temperature $T_0$ corresponding to the mean density and 
(iii) the slope $\gamma$ of the effective 
equation of state of the IGM at a mean redshift $z \simeq 2.5$.
We find that 0.8 $<(T_0/10^4 {\rm K})<$ 2.5 and $1.3<\gamma<2.3$. 
while the constraint obtained on $f$ is $0.020^2<f<0.032^2$. 
A reliable lower bound on $J_{-12}$ can be used to put a lower bound on 
$\Omega_B h^2$, which
can be compared with similar constraints obtained from Big Bang 
Nucleosynthesis (BBN) and CMBR studies. We find that if $J_{-12}>1.2$, 
the lower bound on 
$\Omega_B h^2$ is in violation of the BBN value.
 
\end{abstract}


\section{Introduction}
A significant fraction of the baryons at $z\le5$ 
are found in the form of a diffuse intergalactic
medium (IGM), which is usually probed through the
absorption lines produced by them 
on the spectrum of the distant QSOs. It is 
believed that while the metal line systems (detected
through Mg~{\sc ii} or C~{\sc iv} doublets) seen in the QSO spectra
could be associated with the halos of the intervening luminous galaxies
(Bergeron \& Boisse 1991; Steidel 1993), most of the low neutral
hydrogen column density absorption lines (commonly called as
`Ly$\alpha$' clouds) are due to the low amplitude baryonic
fluctuations in the IGM.

Probing the
baryonic structure formation through Ly$\alpha$ absorption lines has two 
advantages. First, there are large number of absorption lines. 
Typically, one can
observe more than a few hundreds of lines per unit redshift range 
along any one line of
sight. This provides us with a large unbiased dataset, using which the
statistical studies can be performed efficiently. 
The second advantage is that the 
Ly$\alpha$ absorption lines are more 
straightforward to model 
than, say, luminous galaxies. The modelling of galaxies is
complicated by the fact that one has to take into account processes like 
the star formation, radiation feedback and so on; these
processes are not that effective in the IGM, and one can ignore them at 
the first approximation.

The study of the IGM can -- potentially -- 
provide us with information about different aspects of 
the the baryonic structures the universe like 
(i) the mass power spectrum (Croft et
al. 1998; Hui 1999; Croft et al. 1999), 
(ii) the total baryonic density ($\Omega_B$) 
and the total photoionisation rate due to the local
ionising background radiation ($J$)
and (iii) the reionisation
history of the universe (Hui \& Gnedin 1997). 

There
have been various numerical and semi analytical models in the 
literature for the IGM, all of
which are based on the view that the Ly$\alpha$ clouds are small scale
density fluctuations as predicted by the models of
structure formation. The hydrodynamical simulations 
(Bond, Szalay \& Silk 1988; Cen et al. 1994; Zhang, Anninos
\& Norman 1995; Hernquist et al. 1996; Miralda-Escud\'e et al. 1996;
Riediger, Petitjean \& M\"ucket 1998;
Theuns, Leonard \& Efstathiou 1998; Theuns et al. 1998;
Dav\'e et al. 1999) incorporate most
of the ongoing physical processes in the IGM and hence they are necessary for
understanding the evolution of the IGM. However, due to limited numerical
resolution and computing power, they are able to probe only a small box size
(10--20 Mpc). Hence people have tried to complement the
numerical studies with analytic and semi analytic ones (Doroshkevich \&
Shandarin 1977; McGill 1990; Bi 1993; Bi, Ge \& Fang 1995; 
Gnedin \& Hui 1996; Hui, Gnedin \& Zhang 1997; Bi \& Davidsen 1997, 
hereafter BD; Choudhury, Padmanabhan \& Srianand 2000, hereafter Paper I). 
The semi
analytic models do not have problems related to 
limited numerical resolution or box
sizes, and can be used to probe a wide range of parameters. 

The numerical simulations
suggest that {\it most} 
of the Ly$\alpha$ lines arise due to linear or quasi-linear
density fluctuations. 
Therefore one can neglect the highly non-linear baryonic processes (like shock 
heating)
as a first approximation. However, since a simple linear density evolution
cannot produce the saturated Ly$\alpha$ systems, one cannot completely 
ignore the
non-linear effects. The non-linear baryonic density can be calculated from
the linear one using some approximation scheme, like the Zeldovich
approximation (Doroshkevich \& Shandarin 1977; McGill 1990; Hui, Gnedin \&
Zhang 1997), or the lognormal approximation
(Bi 1993; Gnedin \& Hui 1996; BD; Paper I). The neutral fraction is
then 
estimated by considering the equilibrium between the rate of
photoionisation due to background radiation and the rate of
recombination estimated from the temperature defined through the
equation of state. All these models depend on various IGM
parameters such as $\Omega_B$, $J$, 
equation of state and the Jeans length, as well as
the cosmological parameters like $\Omega_m, \Omega_{\Lambda},$ etc.

Besides using these simplifying assumptions, BD realised that it is
sufficient to simulate the IGM in 1D rather than in 3D. 
This increases
the computing power drastically, 
and one can probe large box sizes (hundreds of
Mpc) with high enough resolution. 
BD performed a
detailed study of the evolution of the IGM from $z=2$ to 4. They also compared
their predictions of column density distribution 
with hydrodynamical simulations and observations, and found a good agreement.
In this work, we follow the idea proposed by
BD and carry out semi analytic simulations of the low density IGM. 
The results obtained from such simulations are found to be in quite good 
agreement with various observations (as described in Section 5), thus 
indicating that the lognormal approximation might be a reasonable assumption 
for the low density IGM. 
We extend our studies to probe the parameter space and 
constrain the parameters for a
particular redshift bin using different
statistics obtained from the spectrum. 
Since the recovery of cosmological parameters is not possible with ill
constrained  IGM parameters (for a detailed discussion, see Paper I), 
we concentrate only on 
the parameters related to the IGM at a particular redshift (in this 
case, $z=2.41$).
The parameters are the slope of the equation 
of state ($\gamma$), the temperature corresponding to the mean baryonic 
density or the mean temperature($T_0$) [in this paper, we shall use the 
term `mean temperature' to be equivalent to the temperature corresponding 
to the mean baryonic 
density] and a combination 
of the baryonic density parameter ($\Omega_B$) and the total 
photoionisation rate due to the local ionising 
radiation field ($J$), the combination being $f=(\Omega_B h^2)^2/J_{-12}$, 
where $J_{-12}=J/(10^{-12} {\rm s}^{-1})$.
We find that different statistics are sensitive to different
parameters, and hence they can be used simultaneously to constrain the
parameter space. 

In previous studies involving numerical simulations, 
the parameter $f$ is usually determined by demanding that the simulated mean
transmitted flux match with the observations (Rauch et al. 1997; McDonald et
al. 2000a). Then for a given $f$, 
the constraints on $\gamma$ and $T_0$ are  usually                          
obtained by fitting the lower envelope of the $N_{\rm HI}-b$ scatter plot
(Schaye et al. 2000; Ricotti, Gnedin \& Shull 2000; Bryan \& Machacek 2000; 
McDonald et al. 2000b), 
where $N_{\rm HI}$ is the column density and 
$b$ is the thermal velocity dispersion 
(defined as $b=\sqrt{2 k_{\rm B} T/m_p}~$).
However, 
because of limited 
box size in the hydrodynamical simulations, 
the continuum of the transmitted flux
is not well identified and this introduces errors in the calculation of the 
mean transmitted flux. 
In our approach, we constrain all the three parameters simultaneously using 
all the available transmitted flux statistics, thus utilising 
all the information available in the spectrum.

Section 2 gives the basic structure of the simulation strategy. Although the
basic idea is the same as in BD, we repeat some of the details for
completeness. This section also discusses about the various assumptions used at
different stages. We discuss the various parameters used to model the
simulation in Section 3. 
Section
4 contains a very brief discussion on the various statistical quantities
studied in this paper. The next section contains the results, where we 
compare our simulations with available observational data and constrain 
$f$, $\gamma$ and $T_0$.
Finally, we summarise our conclusions in
section 6.

\section[]{Basic Outline of the Simulation}

We describe the structure of the numerical simulation in this section, 
which is
essentially the same as in BD, for completeness and setting up the notation.

Let $P_{\rm DM}^{(3)}(k)$ denote the linear DM power spectrum in 3D at 
the present epoch $(z=0)$. Then the 
power spectrum for any arbitrary $z$ is given 
by 
\begin{equation}
P_{\rm DM}^{(3)}(k,z)=D^2(z) P_{\rm DM}^{(3)}(k),
\end{equation}
where $D(z)$ gives the evolution of the linear density contrast.
The linear baryonic power spectrum is related to the DM power spectrum 
through the relation (Fang et al. 1993)
\begin{equation}
P_B^{(3)}(k,z)=\f{P_{\rm DM}^{(3)}(k,z)}{(1+x_b^2(z)k^2)^2},
\label{eq:pk_b}
\end{equation}
where
\begin{equation}
x_b(z)=\f{1}{H_0}\left[\f{2 \gamma k_{\rm B} T_m(z)}
{3 \mu m_p \Omega_m (1+z)}\right]
^{1/2}
\label{eq:xb}
\end{equation}
is the Jeans length; $\mu$ is the mean molecular weight of the IGM, 
given by $\mu=4/(8-5Y)$, where $Y$ is the helium weight fraction. 
(This relation 
assumes that the IGM consists mostly of fully ionised hydrogen and helium. 
In this 
paper, we take $Y=0.24$.) 
$T_m$ is the density averaged 
temperature of the IGM and $\gamma$ is the ratio of specific heats. 
$\Omega_m$ is the cosmological density 
parameter. 
Strictly speaking, 
equation (\ref{eq:pk_b}) is valid only for the case where $x_b$ is 
independent of $z$, but it 
is shown by Bi, Borner \& Chu (1992) that equation (\ref{eq:pk_b}) is a 
good approximation for 
$P_B^{(3)}(k,z)$ even when $x_b$ has a redshift dependence.

At this point, it is appropriate to stress some features of 
the parameter $T_m$. 
The obvious interpretation of $T_m$ will be as the mean temperature of the IGM,
$T_0$ (the temperature at the mean density), i.e., $T_m=T_0$. However,
according to BD, using $T_0$ in equation 
(\ref{eq:xb}) 
leads to a value of the 
linear baryonic density fluctuation, $\sigma_B$, larger than
what we expect from hydrodynamical simulations. Hence, they suggested the
use of a density averaged temperature. Since 
$T_m$ appears only in the expression for the Jeans length 
(equation (\ref{eq:xb})), it can also be defined as the effective 
temperature which
determines the Jeans length. 
It is 
clear that the combination  $\gamma T_ m$ 
can, in principle, be fixed if $\sigma_B$ is known through 
hydrodynamical simulations. In this work, we choose 
$\sigma_B(z=2.41)$ to be 1.34, which gives 
$\gamma T_m=5.115\times10^4$K. Our choice of $\sigma_ B(z=2.41)$
is consistent with that of BD (see their Figure 3). Also, 
simulations of Carlberg \& Couchman (1989) give 
$\sigma_B(z=2.8)=0.953$, but one should note that the power spectrum 
they used was normalised to a value that was 
1.4 times smaller than ours. Furthermore, using a 
power spectrum normalised to a value 1.3 times larger than ours, 
Gnedin (1998) obtains 
$\sigma_B(z=2.85)=2.25$. All these values are consistent 
with our choice for which the value of Jeans length at $z=2.41$ is 
$0.12 \Omega_m^{-1/2}$\hmpc. For the background cosmology with $\Omega_m=0.4,
\Omega_{\Lambda}=0.6$, this corresponds to a velocity scale of 
22.3 km s$^{-1}$.
Since we shall be mainly concerned with a small redshift bin
($\Delta z=0.58$), the evolution of $T_m$ should not affect the
results significantly. Hence, we take $T_m$ to be independent of $z$.

Once the power spectrum of linear density perturbations in 3D is obtained, one 
can obtain the corresponding power spectra for density (as well as velocity) 
perturbations in 1D. One can show that the baryonic power spectrum in 1D is 
given by
\be
P_B^{(1)}(k,z)=\f{1}{2 \pi} \int_{|k|}^{\infty} {\rm d}k^{\prime}~k^{\prime}~
P_B^{(3)}(k^{\prime},z).
\e
while the power spectrum for linear velocity perturbations in 1D is 
\be
P_v^{(1)}(k,z)=\dot{a}^2(z) k^2 \f{1}{2 \pi} \int_{|k|}^{\infty} 
\f{{\rm d}k^{\prime}}{k^{\prime 3}} P_B^{(3)}(k^{\prime},z),
\e
where $a$ is the scale factor and $\dot{a}$ is given by the Friedman 
equations
\be
\dot{a}^2(z)=H_0^2 \left[\Omega_m(1+z)+\Omega_k+
\f{\Omega_{\Lambda}}{(1+z)^2}\right],
\e
with
\be
\Omega_k=1-\Omega_m-\Omega_{\Lambda}.
\label{eq:om_k}
\e
The density and the velocity fields are correlated with the correlation being 
given by
\be
P_{\rm Bv}^{(1)}(k,z)={\rm i}~\dot{a}(z) k \f{1}{2 \pi} \int_{|k|}^{\infty} 
\f{{\rm d}k^{\prime}}{k^{\prime}} P_B^{(3)}(k^{\prime},z).
\e

To simulate the density and velocity fields in 1D for a particular redshift 
$z$, we follow the procedure 
given by Bi (1993). We start with two independent Gaussian fields, 
$w_0(k)$ and $u_0(k)$, having unit power spectrum, i.e., 
$\langle w_0^*(k) w_0(p) \rangle=\langle u_0^*(k) u_0(p) \rangle
=2 \pi \delta_{\rm Dirac}(k-p)$. We can then get two independent Gaussian 
fields having power spectra $P_w(k,z)$ and $P_u(k,z)$ respectively 
\be
w(k,z)=w_0(k) \sqrt{P_w(k,z)},~u(k,z)=u_0(k) \sqrt{P_u(k,z)}.
\e
We choose these power spectra to be of the following form (Bi 1993; BD)
\be
P_w(k,z)=\beta^{-1}(k,z)\f{1}{2 \pi} \int_{|k|}^{\infty} 
\f{{\rm d}k^{\prime}}{k^{\prime}} P_B^{(3)}(k^{\prime},z)
\e
and
\be
P_u(k,z)=\f{1}{2 \pi} \int_{|k|}^{\infty} 
{\rm d}k^{\prime}~k^{\prime}~P_B^{(3)}(k^{\prime},z)-P_w(k,z),
\e
where
\be
\beta(k,z)=\f{\int_{|k|}^{\infty} 
({\rm d}k^{\prime}/k^{\prime 3}) P_B^{(3)}(k^{\prime},z)}
{\int_{|k|}^{\infty} 
({\rm d}k^{\prime}/k^{\prime}) P_B^{(3)}(k^{\prime},z)}.
\e
The linear density and the velocity fields in the $k$-space are then given by
\be
\delta_B(k,z)=w(k,z)+u(k,z),
\e
\be
v(k,z)={\rm i}~\dot{a}k\beta(k,z)w(k,z).
\e
The corresponding fields $\delta_B(x,z)$ and $v(x,z)$ in the real 
comoving space 
are obtained by using Fourier transforms. One should keep in mind that the 
above analysis is done in the framework of linear perturbation theory.

However, to study the properties of the IGM one has to take into
account the non-linearities in the density distribution and various
physical processes such as shocks, radiation field, cooling
etc. Detailed hydrodynamical modelling of IGM has shown that most
of the low column density Ly$\alpha$ absorption (i.e. $N_{{\rm HI}}\le10^{14}$
cm$^{-2}$) are produced by regions that are either in the linear 
or in the weakly non-linear regime (Cen et al. 1994; Zhang et al. 1995; 
Hernquist et al. 1996; Miralda-Escud\'e et al. 1996; 
Theuns, Leonard \& Efstathiou 1998; Theuns et al. 1998; 
Dav\'e et al. 1999). The lower envelope of the column density, 
$N_{\rm HI}-b$ 
scatter plot  
(Schaye et al. 1999; Schaye et al. 2000) 
suggests that 
there is a well defined relationship between the density and the
temperature of the IGM (Hui \& Gnedin 1997). Thus it is
possible to model low column density systems using simple prescription for
the non-linear density field and an equation of state.

Following BD, we take into account the effect of non-linearities of density 
perturbations by
assuming the number density distribution of the baryons, 
$n_B(x,z)$ to be
a lognormal random field
\be
n_B(x,z)=A~{\rm e}^{\delta_B(x,z)}
\e
where $\delta_B(x,z)$ is the linear density contrast in baryons, and 
$A$ is a constant to be determined. The mean value of 
$n_B(x,z)$ is given by
\be
\langle n_B(x,z) \rangle \equiv n_0(z)
= A \langle {\rm e}^{\delta_B(x,z)} \rangle,
\e
where $n_0(z)$ is related to the baryonic density parameter 
$\Omega_B$ through the relation 
\be
n_0(z)=\f{\Omega_B \rho_c}{\mu_B m_p} (1+z)^3.
\e
Here $\rho_c=1.8791\times10^{-29}h^2~$cm$^{-3}$ is the critical density 
of the universe and $\mu_B m_p$ is the mass per baryonic particle, 
given by $\mu_B m_p=4m_p/(4-3Y)$.
Hence, we get the value of the constant as
\be
A=\f{n_0(z)}{\langle {\rm e}^{\delta_B(x,z)} \rangle}
\e
and
\be 
n_B(x,z)=n_0(z) \f{{\rm e}^{\delta_B(x,z)}}
{\langle {\rm e}^{\delta_B(x,z)} \rangle}.
\label{eq:ln}
\e
The lognormal distribution was introduced by 
Coles \& Jones (1991) as a model for the non-linear matter distribution in the 
universe. Detailed arguments as to why this ansatz should be 
reasonable in studying non-linear density distribution 
can be found in 
Coles \& Jones (1991), BD and Paper~I. 
However, we would still like to 
stress some points regarding the use of the lognormal ansatz for baryons in 
the current context.

In the past, there have been attempts to use the lognormal 
distribution to model the
dark matter. However, we now know (based on Non-linear Scaling Relations; see
Nityananda \& Padmanabhan 1994; Padmanabhan 1996) 
that any local mapping of the form 
$\delta_{\rm NL}~=~F[\delta_{\rm L}]$ is
bad for dark matter (also see Coles, Melott \& Shandarin 1993).
There is, however, a strong theoretical argument (see
Paper~I) which shows that lognormal produces the
correct limits at the two extremes for {\it baryons.}  
At large spatial scales, where the
density contrast is small ($\delta_{\rm B} \ll 1$), 
equation (\ref{eq:ln}) reduces 
to 
$n_{\rm B}/n_0 \simeq 1+\delta_{\rm B}$, which is just what 
we expect from linear theory.
More importantly, on small 
scales, equation (\ref{eq:ln}) becomes the isothermal hydrostatic solution, 
which 
describes highly clumped structures like intracluster gas, 
$n_{\rm B} \propto \exp 
(-\mu m_p \psi_{\rm DM} /\gamma k_B T)$, where 
$\psi_{\rm DM}$ is the dark matter
potential (Sarazin \& Bahcall 1977). This gives an indication that even though 
the lognormal ansatz is poor for dark matter distribution, it might 
still work reasonably well for baryonic matter since it is 
correctly constrained at both extremes.

Comparison with full hydrodynamical simulations reinforces this conclusion. 
BD have used the results from the hydrodynamical 
simulations of Miralda-Escud\'e et al. (1996), 
and found that the baryonic density distribution can be 
well fitted by a lognormal function at $z=3$.  
Also, the range of parameters for the IGM 
recovered by us and those by the
full hydrodynamical simulations of McDonald et al. (2000a, b)
agree quite well (as discussed later in Section 6). 
This shows that the lognormal
assumption agrees with the hydrodynamical simulations as well.

The above arguments should convince the reader that the lognormal 
assumption, in spite of its limitations, provides us with a 
tool in studying the baryonic structure formation semi analytically.
No such approximation can reproduce the
results obtained from the full hydrodynamical simulations 
exactly and their values lie in providing faster route to
reasonably accurate results. Our attempts should be viewed in the backdrop
of such a philosophy. 

There is no obvious way to deal with the non-linearities in the velocity
field, but fortunately this is not needed in the current work as can be seen 
from the following argument. 
The velocity field plays two separate roles in the context of our work. 
The first one is that 
the velocity determines the movement of the individual particles at a 
given instant of time, 
which in turn affects the underlying density field in the next time step. 
Mathematically this feature is represented by the Euler equation, 
which connects the density field to velocity field. 
Given any prescription for density field, the Euler equation implicitly 
leads to a consistent velocity field. Hence
this dynamical effect of velocity field -- viz. moving the mass to 
the right location -- is indirectly
taken into account in any prescription for non-linear density. 
In our case, the 
lognormal ansatz takes care of this feature.
The second effect of the velocity field is purely kinematic -- it 
shifts the positions of the 
absorption lines. 
In our work we will be using the thermal velocity dispersion $b$ 
in the Voigt profile while analysing the lines. Since the 
Ly$\alpha$ absorption lines originate from quasi-linear density regions, 
the velocity field 
will be subdominant or 
of the same order as the thermal velocity dispersion $b$, which will be 
taken care of in our analysis. 

Once the non-linear baryon density is obtained, it is trivial to get the 
fraction of hydrogen in the neutral form, $f_{\rm HI}$,
in the IGM by solving the ionisation
equilibrium equation for hydrogen
\be
\alpha(T) n_p n_e=\Gamma_{\rm ci}(T) n_e n_{\rm HI} + J n_{\rm HI},
\label{eq:ion_eq}
\e
where $\alpha(T)$ is the radiative recombination rate, $\Gamma_{\rm ci}(T)$ 
is the rate of collisional ionisation and $J$ 
is rate of photoionisation for hydrogen (Black 1981); $n_p, 
n_e$ and
$n_{\rm HI}$ are the number densities of proton, 
electron and neutral hydrogen, 
respectively. In general, all these quantities are functions of $z$ and all 
except $J$ depend on the position $x$ too. 
We shall parametrise $J(z)$ by a dimensionless quantity 
$J_{-12}(z)$, defined by $J(z)=J_{-12}(z) 10^{-12}$s$^{-1}$. For comparison, 
we mention that our $J$ is equal to the quantity $J_{21} G_1$ used by BD.

Black (1981) gives the approximate form of the recombination and ionisation 
rates as follows: 
\be
\f{\alpha(T)}{{\rm cm}^3{\rm s}^{-1}}=\left\{ \begin{array}{ll}
	    4.36\times10^{-10}T^{-0.7573}
				&\mbox{(if $T \geq 5000$K)}\\
	    2.17\times10^{-10}T^{-0.6756}
				&\mbox{(if $T < 5000$K)}
		\end{array} \right.
\e
and
\be
\Gamma_{\rm ci}(T)=
5.85\times10^{-11} T^{1/2} \exp(-157809.1/T)~{\rm cm}^3{\rm s}^{-1},
\e
where $T$ is in Kelvin.
One can see that the expression for $\alpha(T)$ diverges as 
$T \to 0$ which needs to be regularised by  
a temperature cutoff at the lower end in
numerical work. BD have 
used the photoionisation temperature as the minimum temperature, which is 
about $10^4$K (see also Theuns et al. 1998). 
In situations where $T_0>10^4$K, we too 
shall use the same value. However, when $T_0 \le 10^4$K, we have taken the 
minimum temperature to be $5000$K.

The IGM contains mainly hydrogen, a smaller amount of helium
(weight fraction, $Y\sim 0.24$) 
and negligible amount of other heavier elements. In that case 
we can write $n_e=\kappa n_p$, where $\kappa$ is a constant, greater than but 
very close to unity. However, in the following calculation we have 
neglected the 
presence of the heavier elements completely for simplicity.
Let us define the neutral fraction of hydrogen, 
$f_{\rm HI}$ by
\be
f_{\rm HI}=\f{n_{\rm HI}}{n_B}=\f{n_{\rm HI}}{n_{{\rm HI}}+n_p} 
\label{eq:f_def}
\e
(we ignore the number density contributed be helium because, usually, 
$n_{\rm He}/n_B < 0.1$).
Using equation (\ref{eq:ion_eq}) in (\ref{eq:f_def}), one gets
\be
f_{\rm HI}(x,z)=
\f{\alpha(T(x,z))}{\alpha(T(x,z))+\Gamma_{\rm ci}(T(x,z))+J(z)/n_e(x,z)}.
\e
We express $n_e$ in terms of $n_B$ by assuming that $f_{\rm HI} \ll 1$ and 
all the helium 
present is in the fully ionised form. In such case,
\be
n_e/n_B\equiv\mu_e=2(2-Y)/(4-3Y).
\e 
Then,
\be
n_{\rm HI}(x,z)=\f{\alpha(T(x,z)) n_B(x,z)}
{\alpha(T(x,z))+\Gamma_{\rm ci}(T(x,z))+J(z)/(\mu_e n_B(x,z))}
\label{eq:nh_nb}
\e
We calculate $T(x,z)$ using the polytropic equation of state 
\be
T(x,z)=T_0(z) \left(\f{n_B(x,z)} {n_0(z)}\right)^{\gamma-1},
\e
where $T_0(z)$ is the temperature of the IGM at the mean density. 
We now know the neutral hydrogen density at any redshift $z$ along a 
particular
axis. Our next goal is to find the density along a line of sight. This 
can be done by obtaining the density field along the backward 
light cone. In other 
words, we must obtain the quantity $n_{\rm HI}(x,z(x))$, where $x$ and $z$ 
are related through the expression
\be
x(z)=\int_0^z d_H(z^{\prime})~{\rm d}z^{\prime},
\label{eq:x(z)}
\e
with
\begin{eqnarray}
d_H(z)&=&c\left(\f{\dot{a}}{a}\right)^{-1} \nonumber \\
      &=&\f{c}{H_0} [\Omega_{\Lambda}+\Omega_m (1+z)^3+\Omega_k
(1+z)^2]^{-1/2}.
\label{eq:d_h}
\end{eqnarray}
Similarly one can also get the velocity field $v(x,z(x))$ along the same LOS. 
Once the neutral hydrogen density and the velocity along the LOS is known, 
the Ly$\alpha$ absorption optical depth at 
redshift $z_0$ can be obtained from the relation
\begin{eqnarray}
\tau(z_0)&=&\f{c I_{\alpha}} {\sqrt{\pi}} \int~{\rm d}x
\f{n_{\rm HI}(x,z(x))}{b(x,z(x))(1+z(x))} \nonumber \\ 
&\times& V\left[\alpha,\f{c(z(x)-z_0)}{b(x,z(x))(1+z_0)}+
\f{v(x,z(x))}{b(x,z(x))}\right],
\end{eqnarray}
where 
\be
b(x,z(x))=\sqrt{\f{2k_{\rm B}T(x,z(x))}{m_p}},
\e 
$I_{\alpha}=4.45\times10^{-18}$cm${^2}$ and 
$V$ is the Voigt function. $I_{\alpha}$ is related to the 
Ly$\alpha$ absorption cross section through 
\be
\sigma_{\alpha}(\nu)=\f{c I_{\alpha}}{b \sqrt{\pi}}~ 
V\left[\alpha,\f{c(\nu-\nu_{\alpha})}{b \nu_{\alpha}}\right].
\e 
For low column density 
regions, the natural broadening is 
not that important, and the Voigt function reduces to 
a simple Gaussian
\be
V[\alpha,\f{\Delta v}{b}]\simeq\exp\left(-\f{(\Delta v)^2}{b^2}\right).
\e
Since we are mostly dealing with weakly non-linear regimes, where the 
densities are not too high, this 
approximation does not introduce any significant error in the final results.
The optical depths obtained above are used to get the final line of
sight spectrum.

\section{Model Parameters}

Following Paper I, the
model parameters can be broadly divided into two classes, namely, those 
related to the background cosmology and those related to the baryonic IGM. 
In Paper I, we 
have considered various CDM cosmological models (SCDM, LCDM, OCDM) and 
a range of IGM parameters, and have found that the parameters 
related to the background cosmology cannot be 
constrained uniquely with ill defined IGM parameters. Consequently, 
the approach taken in this paper is to use the available observations
for constraining the IGM related
parameters under the framework of most favoured structure formation
scenario. We consider the following parameters for the LCDM model. 
The CDM power spectrum in 3D is taken to be (Efstathiou, Bond \& 
White 1992)
\be
P_{\rm DM}^{(3)}(k)=\f{A_{\rm DM} k}{(1 + [a_1 k + (a_2 k)^{1.5} + 
(a_3 k)^{2}]^{\nu})^{2/\nu}}
\e
where $\nu=1.13$, 
$a_1=(6.4/\Gamma)$\hmpc, $a_2=(3.0/\Gamma)$\hmpc, $a_3=(1.7/\Gamma)$\hmpc\ and 
$\Gamma=\Omega_m h$. The 
normalisation parameter $A_{\rm DM}$ 
is fixed through the value of $\sigma_8$ (the 
rms density fluctuation in spheres of radius 8 \hmpc) which is taken to be 
$\sigma_8=0.79$.
The other model parameters are: 
\be
\Omega_m=0.4, \Omega_{\Lambda}=0.6, h=0.65.
\e
The values of $\Omega_m$ and $\Omega_{\Lambda}$ are consistent with the 
best fitted parameters of Ostriker \& Steinhardt (1995). The value of 
$\sigma_8$ is obtained from the first year COBE normalisation (Kofman, 
Gnedin \& Bahcall 1993). This value is also consistent with those obtained 
from the observed local abundance of clusters by Eke, Cole \& Frenk (1996). 
An identical LCDM model is considered in the hydrodynamical 
simulations by Miralda-Escud\'e et al. (1996). 

Once the cosmology is fixed, we turn our attention towards the parameters 
related to the baryons.
\begin{enumerate}
\item{Slope of the effective equation of state (${\gamma}$): }
It is known that
the value of $\gamma$, at any given epoch, depends on the reionisation history
of the universe (Theuns et al. 1998, Hui \& Gnedin 1997). 
The value of $\gamma$ and its evolution 
are still quite uncertain. Using
Voigt profile fits to the observed Ly$\alpha$ absorption lines one can in
principle obtain the value of $\gamma$.  
In this work, we will keep $\gamma$ as a free parameter 
and ignore its redshift evolution. 

\item{Mean temperature ($T_0(z)$): }
Mean temperature of the IGM is decided by the various heating and
cooling processes. In addition to the reionisation history local
radiation field will also affect the value of $T_0(z)$. In the case of
full hydrodynamical models, the mean temperature is estimated
self-consistently by considering various processes. However in our
approach we consider the mean temperature as a free parameter. 
We also take it as independent of $z$ within the small
redshift bin we consider.

\item{${\Omega_B h^2}$ and $J_{-12}(z)$: }
If one compares the typical values of the three quantities in the 
denominator in the right hand side 
of equation (\ref{eq:nh_nb}), one can verify that $\alpha(T)$ 
and $\Gamma_{\rm ci}(T)$ are much smaller compared to $J/n_B$ (typically, 
for $T \sim 5\times10^4$K, $\alpha \sim 10^{-13}$cm$^3$s$^{-1}$, 
$\Gamma_{\rm ci} \sim 10^{-14}$cm$^3$s$^{-1}$ and 
$J(z)/n_B \sim 10^{-5}$cm$^3$s$^{-1}$). This means that we can 
write $n_{\rm HI} \simeq \alpha n_B^2/J$. Since 
$n_B \propto n_0 \propto \Omega_B h^2$ and 
$J(z) \propto J_{-12}(z)$, we see that only the 
combination $f(z)=(\Omega_B h^2)^2/J_{-12}(z)$ 
appears in the expression for optical depth. We shall treat this 
quantity $f(z)$ as a free parameter. We shall also assume that the 
photoionisation rate does not depend on $z$ (at the least, it does not vary 
considerably within the small redshift bin we are interested in). This will 
make $f$ independent of $z$.

\end{enumerate}

So, we finally end up with three free parameters, namely $\gamma$, $T_0$, 
and $f$.

\section{Statistical Quantities: Definitions}

We perform various statistics on our simulated spectrum, as one usually 
does with the real data, to constrain various parameters of our model.

>From the spectrum, one can immediately calculate the mean transmitted flux
($\bar{F}$) and the rms flux fluctuations ($\sigma_F^2$). 
The transmitted flux data can also be used to obtain three important 
statistics (McDonald et al. 2000a). These are: 
(i) the probability distribution function (PDF) for the transmitted flux, 
(ii) the correlation function of the transmitted flux, 
defined as 
$\xi(\Delta v)=\langle (F(v)-\bar{F})(F(v+\Delta v)-\bar{F}) \rangle$ and 
(iii) the flux power spectrum $(P_F(k))$. 
The power spectrum is calculated using the Lomb periodogram technique (Lomb
1976; Scargle 1982; Press et al. 1992) and the normalisation used is the same 
as mentioned in McDonald et al. (2000a), i.e., 
$\sigma_F^2=\int_{-\infty}^{\infty}({\rm d}k/2\pi)P_F(k)$. 
The advantage in the case of statistics obtained from transmitted flux is that
the numerical 
procedure is quite fast. We use these statistics for constraining the 
parameter space. 

For a set of most favourable values of parameters, we decompose 
the spectrum into individual lines using Voigt profile
analysis, and use them to check our predictions with the observations. 
The statistics used for this purpose are (i) the number of lines (absorbers) 
per unit redshift range (${\rm d}N/{\rm d}z$) and the
column density distribution ($f(N_{\rm HI})$), defined as the number of 
lines (absorbers) per
unit redshift path per unit column density range (Kim et al. 1997), (ii) the 
distribution of the $b$ parameter  and 
(iii) the two point correlation function for the absorbers, defined as 
$\xi_{\rm cloud}(\Delta v)=[N_{\rm sim}(\Delta v)/N_{\rm exp}(\Delta v)]-1$, 
where $N_{\rm sim}(\Delta v)$ is the number of 
cloud pairs with a velocity separation $\Delta v$ 
obtained from the simulated data and $N_{\rm obs}(\Delta v)$ is 
the number of pairs expected from a random distribution of clouds 
(Sargent et al. 1980;
Webb 1987; Srianand \& Khare 1994; Kulkarni et al. 1996; Srianand 1996; 
Khare et al. 1997; Kim et al. 1997; Cristiani et al. 1997).

\section{Results}

In this paper, we have concentrated our studies in the redshift range 
2.09--2.67. This range corresponds to a box size of 436 $h^{-1}$ Mpc 
(for the cosmology we are considering), which is virtually 
impossible to probe in a full 3D hydrodynamical simulation with 
high enough resolution. 
The number 
of grid points used in this work along the line of sight 
was $2^{15}$, which were equispaced in the 
comoving coordinate $x$.
The simulated flux data was then 
resampled with $\Delta \lambda$=0.04 \AA\ and a random noise 
of (S/N)=30 was added, exactly as is done with the 
observed data. We mention here that even if we
increase the number of points (i.e., try to achieve a better resolution),
the resampling mentioned above would make sure that the statistics obtained 
from the transmitted flux are not affected.
We found that the continuum of the spectrum is quite well 
defined at this redshift, and hence it was not necessary to 
make any extra normalisation.

\subsection{Comparison with observations}

We use
various statistics obtained from the observational data given 
by McDonald et al (2000a). In the redshift range 2.09--2.67, they have 
considered data from 5 QSOs, namely, Q2343+123 ($z_{\rm em}=2.52$), 
Q1442+293 ($z_{\rm em}=2.67$), 
KP77:~1623+2653 ($z_{\rm em}=2.526$), 
Q1107+485 ($z_{\rm em}=3.00$) and 
Q1425+604 ($z_{\rm em}=3.20$). 
Each of these quasar sight lines span different regions of 
the redshift interval, and hence 
all the redshifts are not
equally weighted in the above mentioned redshift range. However
in our simulated data we cover the same redshift range giving
equal weightage. 
We have 
confirmed that the correction introduced due to this uneven 
weightage in observed data is negligible (i.e., much below 
typical observational errors).  

\begin{figure*}
\psfig{figure=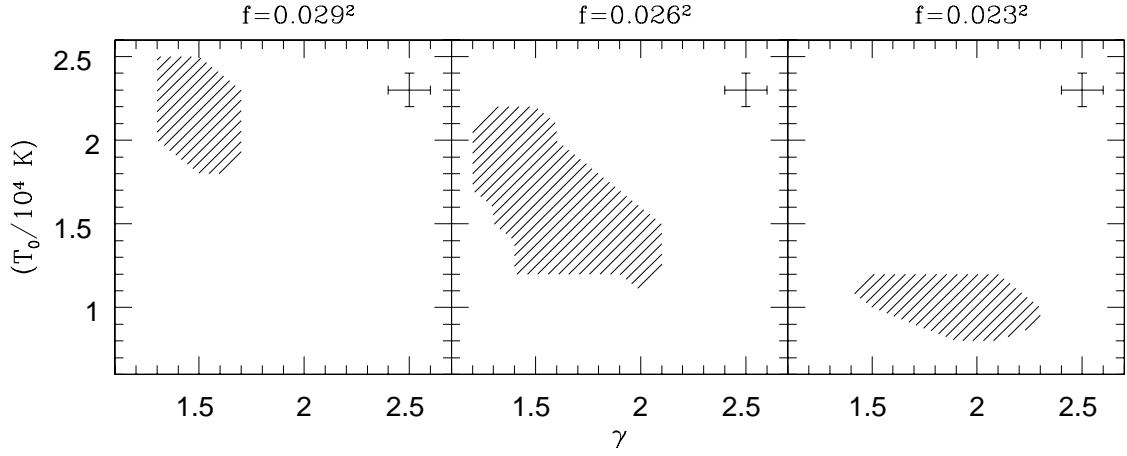,width=16cm}
\caption[]{\label{t0_gamma}
The constraints obtained in the $\gamma-T_0$ space for different values of 
$f$, {\it using transmitted flux statistics}. The shaded regions 
denote the range allowed by observations. The boundaries are 
uncertain by an amount 0.1 along $\gamma$ axis and by 1000K along the 
$T_0$ axis because of finite sampling, which is shown by a cross at the upper
right hand corners of the panels.
}
\end{figure*}

\begin{figure*}
\psfig{figure=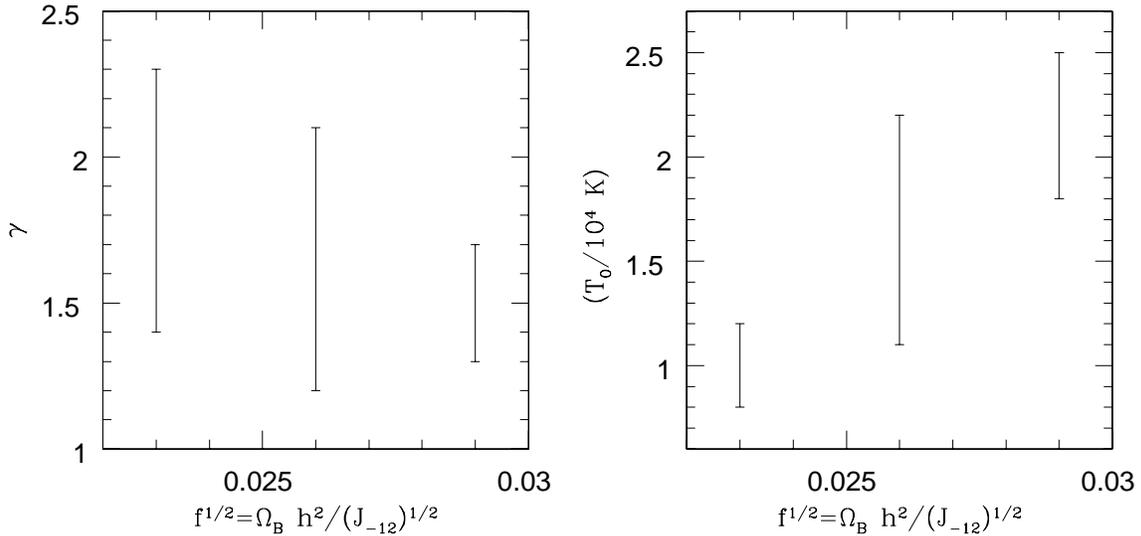,width=16cm}
\caption[]{\label{allowed}
The left panel shows the allowed range of $\gamma$ for different values of 
$f$, regardless of the value of $T_0$. The right panel shows the allowed 
range of $T_0$, regardless of $\gamma$. The ranges are obtained using 
transmitted flux statistics only.
}
\end{figure*} 

The allowed range for various parameters are
obtained by demanding
that the simulated data pass through most of the observed points, within
the allowed 1$\sigma$ error limits.
The value of $f$ is strongly constrained between $0.020^2$--$0.032^2$ 
(regardless of the value of $\gamma$ and $T_0$). 
In the above range of $f$, we consider models with $f=0.023^2,
0.026^2,0.029^2$ and 
obtain constrains on $T_0$ and $\gamma$ so that all
the observed statistics obtained from the transmitted flux 
are consistently reproduced.

The constrained parameter space for these three values 
of $f$ is shown in Figure \ref{t0_gamma}. (We mention here that due to 
finite sampling of the parameter space, the boundaries of the allowed region 
are uncertain by an amount 0.1 along the $\gamma$ axis and by 1000K along the 
$T_0$ axis; this error budget is 
indicated by a cross at the right top corner of the panels.)
It is obvious that as we go to lower values of $f$, the observations allow 
lower values of $T_0$ and higher values of $\gamma$. 
For example, $1.4<\gamma<2.3$, $0.8 \times 10^4$K$<T_0<1.2\times 10^4$K 
for $f=0.023^2$ whereas 
$1.3<\gamma<1.7$, $1.8\times 10^4$K$<T_0<2.5\times 10^4$K for $f=0.029^2$.
It is also seen that the area of the allowed region is maximum for $f=0.026^2$ 
and is smaller for higher or lower values of $f$. We mention here that 
the allowed region is practically zero for $f<0.020^2$ and $f>0.032^2$.

The limits on $\gamma$ and $T_0$ for $f$ in the range $0.023^2$--$0.029^2$ 
are shown in Figure \ref{allowed}. 
The left panel shows the allowed range of $\gamma$ 
regardless of the value of $T_0$, the right panel shows that for 
$T_0$, regardless of $\gamma$. 

Although the observations allow $f$ in the range $0.020^2$ to $0.032^2$, 
we find that the match between simulations and observations is best for 
$f\sim 0.026^2$. We calculated the $\chi^2$ of the three 
statistics for different parameter values and found that they are 
comparatively lower for $f=0.026^2$ than for higher or lower values of 
$f$. Hence, 
in what follows, we shall concentrate on $f=0.026^2$, and see explicitly 
how all the statistics compare with 
observations.
\begin{figure}
\psfig{figure=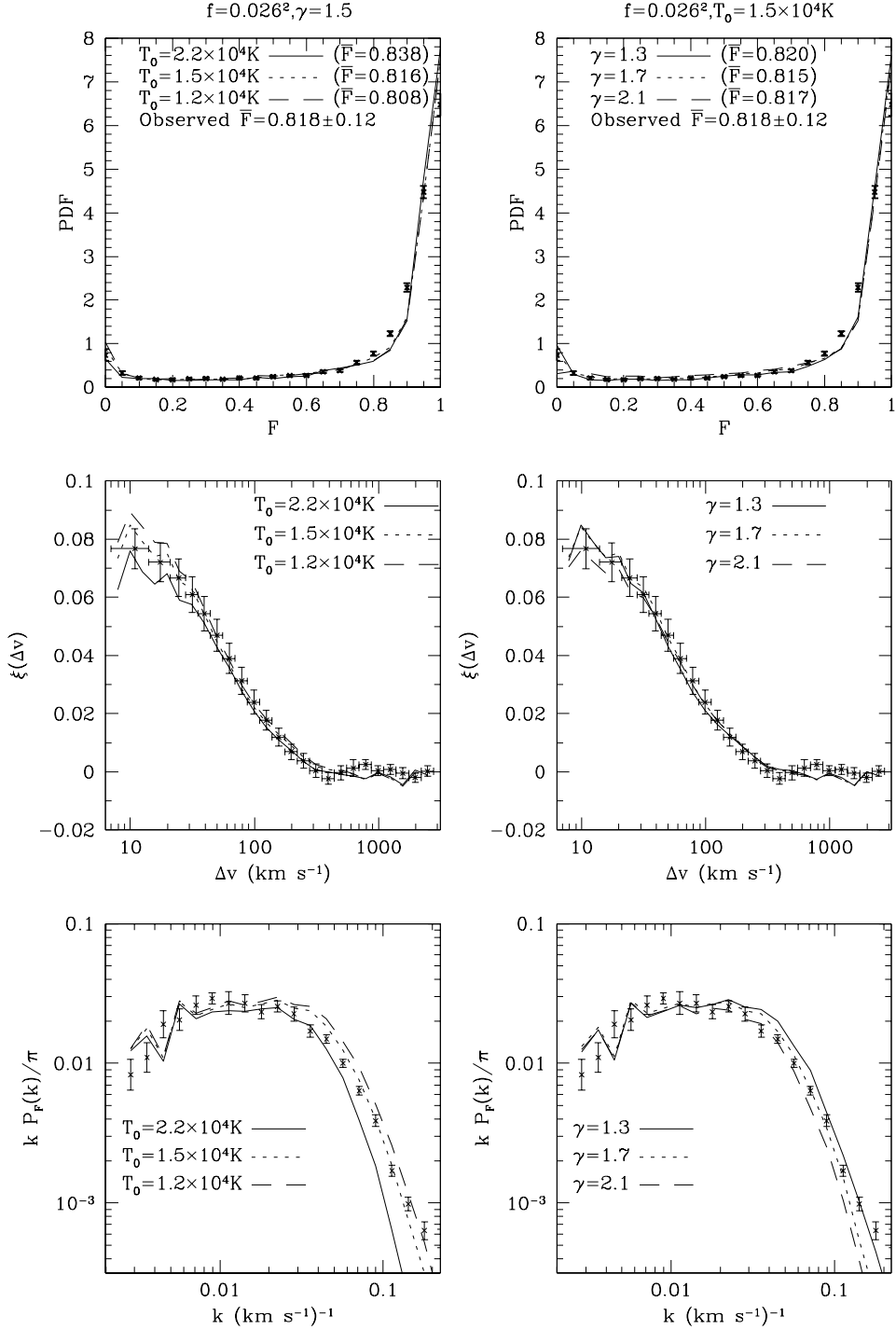,width=14cm}
\caption[]{\label{stat_allowed}
The comparison between simulations and observed results for $f=0.026^2$. The 
points with errorbars are the observed data points (McDonald et al. 2000a). 
In the left panel, we 
show the limits on $T_0$ for a particular $\gamma$ (in this case, 1.5), 
and the right panel shows the limit on $\gamma$ for $T_0=1.5 \times 
10^4$K. 
}
\end{figure}
In Figure \ref{stat_allowed}, we show the comparison between simulations 
and observations for various transmitted flux statistics for 
some particular values of $\gamma$ and $T_0$. 
The point to be noted here is that {\it our simulations are able to 
match the observations for all the three statistics for a particular range 
of parameter values.} The results obtained using hydrodynamical simulations 
fail to match the observations for the flux correlation 
function and the power spectrum simultaneously (McDonald et al. 2000a), 
mainly because of 
the lack of power at large scales (due to limited 
box size). As discussed earlier, 
our semi analytic simulations probe large box sizes
without compromising on the resolution, and hence we 
are able to match both the statistics simultaneously. We mention here 
that the the typical length scales probed by both the correlation 
function and the power spectrum is about 100 $h^{-1}$ Kpc to 25 $h^{-1}$ Mpc.

The left panel 
shows the limit on $T_0$ for $\gamma=1.5$. It is clear that for 
$T_0\ge2.2 \times 10^4$K, the value of $\bar{F}$ is larger than 
what is allowed by the observations. At temperatures 
higher than this, the recombination rate is so low that the neutral fraction 
of hydrogen reduces and hence the transmitted flux goes above the 
allowed limit. Furthermore, we can see that the power spectrum 
also restricts the allowed range of $T_0$ between 
(1.2--2.2)$\times 10^4$K. At higher (lower)
temperatures, the power at smaller scales are reduced (enhanced)
due to excess (less) Voigt profile smoothing. 
>From the way 
it is defined (see the previous section), 
the normalisation of the correlation function depends on $T_0$. 
The correlation curves go up when $T_0$ is decreased. However, since the
errorbars on $\xi$ are comparatively larger, the correlation 
curve does not impose any further constraints.

The right panel 
shows the limit on $\gamma$ for $T_0=1.5 \times 10^4$K. 
(We mention here again that while 
changing the value of $\gamma$, we change the value of $T_m$ also, such that 
the combination $\gamma T_m$, and hence the Jeans length remains unchanged.) 
The effect of increasing $\gamma$ is to increase the range of
temperature in the IGM for a given baryon density range (Paper I). 
This actually reduces the range in the recombination rate
and hence the range of neutral hydrogen density.  
Since a large $\gamma$ means less fluctuations in the
neutral hydrogen densities, there are
less number points  having
extreme values of the flux, as one can see from the PDF. 
It is clear that one 
can rule out $\gamma>2.1$ from the PDF.
Also, one can see from the flux power spectrum curve 
that 
there is a reduction (enhancement) in the small scale power for 
larger (smaller) values of $\gamma$. This restricts $\gamma$ between 
1.3--2.1.
The correlation curve 
is quite insensitive to $\gamma$ as compared to the other 
two parameters, 
and hence it does not impose any further constraints. 
Since  
the normalisation of the correlation function depends on $f$ and $T_0$ but 
not on $\gamma$, the correlation curves are 
comparatively less sensitive to $\gamma$. 

In the allowed ranges of parameters, the match between observations and 
our simulations is 
quite good for all the three statistics obtained from the transmitted flux. 
We did not compare the simulated flux power
spectrum with observations for smaller scales ($k>0.2$ km s$^{-1}$).
The reason for this is the presence of
narrow metal lines in the observed spectra, which contribute to the small
scale power. Detailed discussion regarding this aspect can be found in 
McDonald et al. (2000a). 

\begin{figure}
\psfig{figure=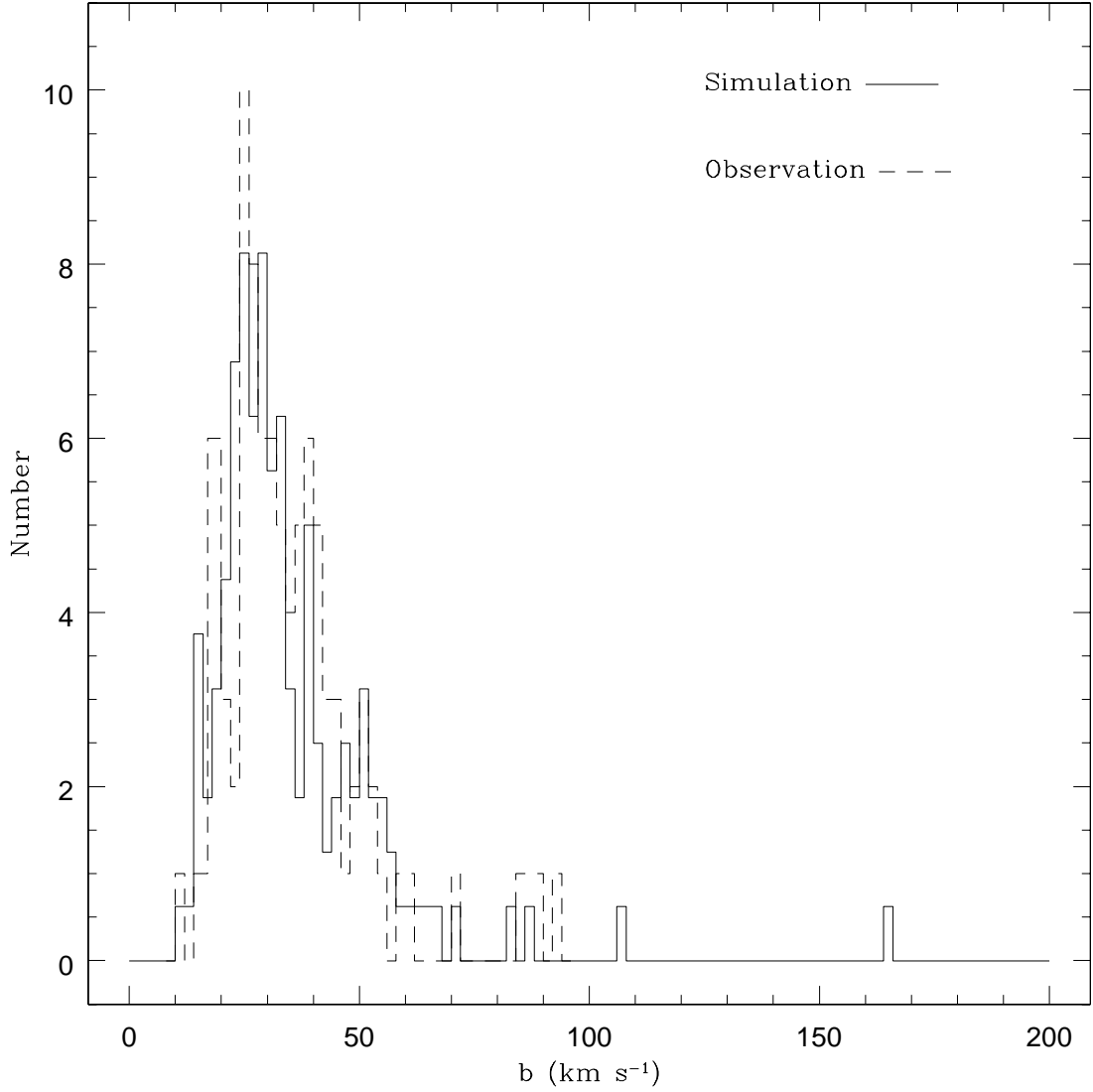,width=16cm}
\caption[]{\label{bdist}
Comparison of the observed $b$ distribution with the simulations. Here, 
$\gamma=1.7$, $T_m=3.01 \times 10^4$K, $f=0.026^2$ and $T_0=1.5 \times 10^4$K. 
The observed data is for $z=2.31$, taken from Kim et al. (1997).
}
\end{figure}

Once we have constrained the range of $\gamma-T_0$ space 
for $f=0.026^2$, 
it is worth checking whether we can
match the observed statistics obtained from the Voigt profile decomposition of
the spectrum. We have used the standard Voigt profile routine 
(Khare et al. 1997) to decompose the observed spectrum into clouds. 
The minimum number of components required to fit an absorption line 
is constrained by the $\chi^2$ minimisation. 
For this purpose, we concentrate on a particular value of 
$T_0=1.5 \times 10^4$K and $\gamma=1.7$. 
For obtaining the statistics, we take
lines centered around $z=2.26$, so as to mimic the observed data of
Kim et al. (1997) at $z=2.31$. The redshift interval considered is 
$\Delta z=0.26$.
Figure \ref{bdist} shows the
comparison between
observations
and simulations for the $b$ distribution. The mean $b$ of the
simulated distribution is 35.19 km s$^{-1}$, whereas that of the observed
distribution is 36.35 km s$^{-1}$. We have performed a $\chi^2$
statistics for the two distributions, and found $\chi^2/\nu=0.61$ (with 
$\nu=34$, 96.3 per cent likelihood).

\begin{figure}
\psfig{figure=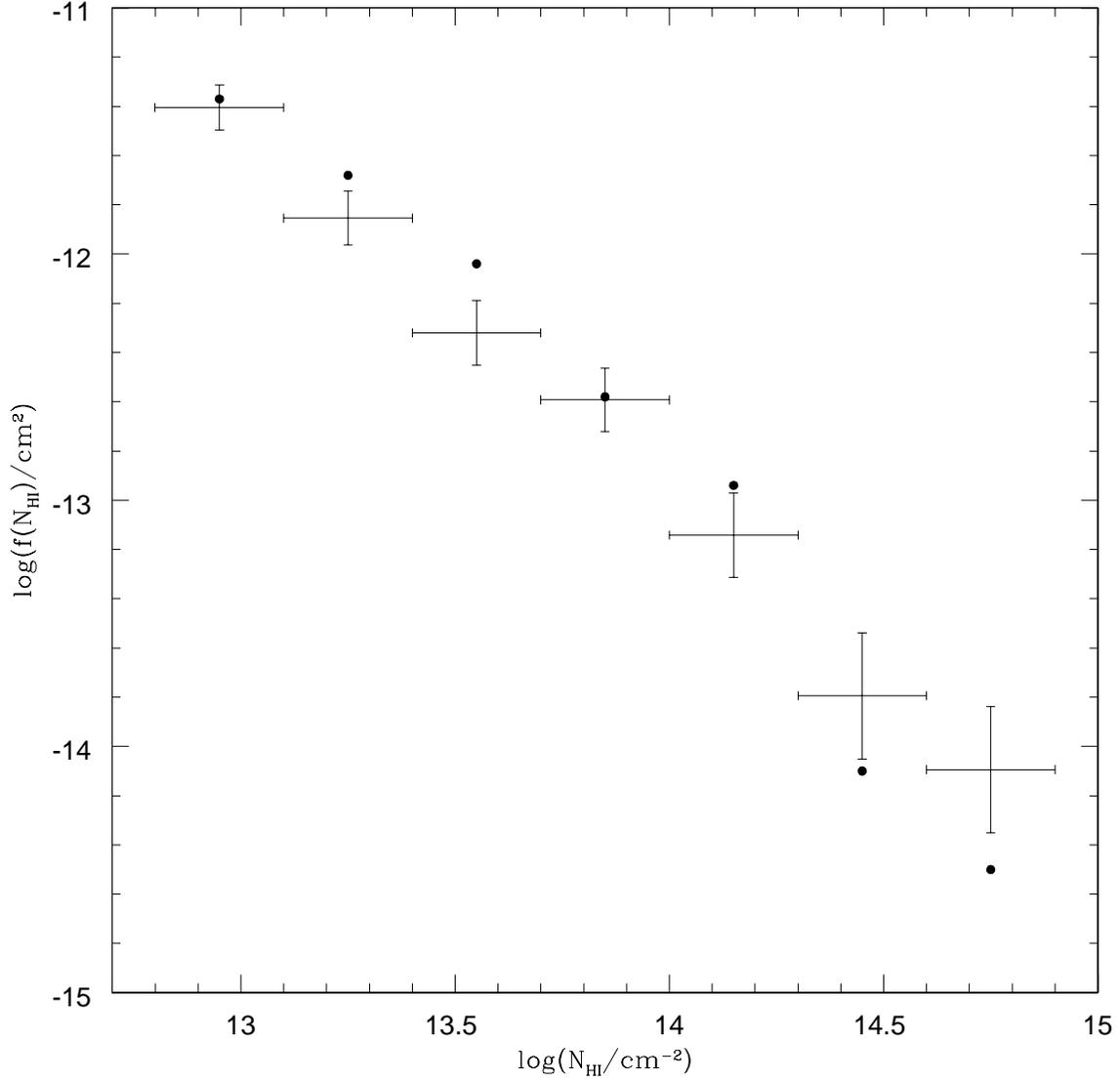,width=16cm}
\caption[]{\label{fnh_dist}
Comparison of the observed column density distribution with the simulations. 
The points in the figure are the observed data. The error bars indicate the
values obtained from simulations. Here,
$\gamma=1.7$, $T_m=3.01 \times 10^4$K, $f=0.026^2$  and $T_0=1.5 \times 10^4$K.
}
\end{figure}

We perform the same exercise with the column density distribution
$f(N_{HI})$. The comparison between simulations and observations is shown in
Figure \ref{fnh_dist}. 
Usually, one assumes a power law distribution for 
$f(N_{HI})$, i.e., $f(N_{HI})
\propto N_{HI}^{-\beta_{HI}}$. In our case, we obtain the slope by 
carrying out a maximum likelihood analysis (Srianand \& Khare 1996) where 
effects of binning are avoided. We present the distribution in the 
figure in the binned form for the purpose of better visualisation.
The slope of the column density distribution,
$\beta_{HI}$, in the column density
range $12.8\le\log(N_{HI}/{\rm cm}^{-2})\le16.0$ obtained from simulations is
$1.31\pm 0.13$; the corresponding quantity obtained from observations is 
around 1.35 (Kim et al. 1997). One can also compare the values of 
${\rm d}N/{\rm d}z$ obtained from simulations and observations. For 
$13.77\le\log(N_{\rm HI}/{\rm cm}^{-2})\le16.0$, we get 
${\rm d}N/{\rm d}z=141.06 \pm 23.19$.
The corresponding number obtained from observations is between 63.09 and 100.00
(see Figure 2 of Kim et al. 1997). 
For 
$13.1\le\log(N_{\rm HI}/{\rm cm}^{-2})\le14.0$, we obtain 
${\rm d}N/{\rm d}z=202.06 \pm 27.75$, 
which is well within the observed limits of 
158.49 and 223.87. 

\begin{figure*}
\setlength{\unitlength}{1cm}
\centering
\begin{picture}(12,10)
\put(-3.5, 0){\includegraphics{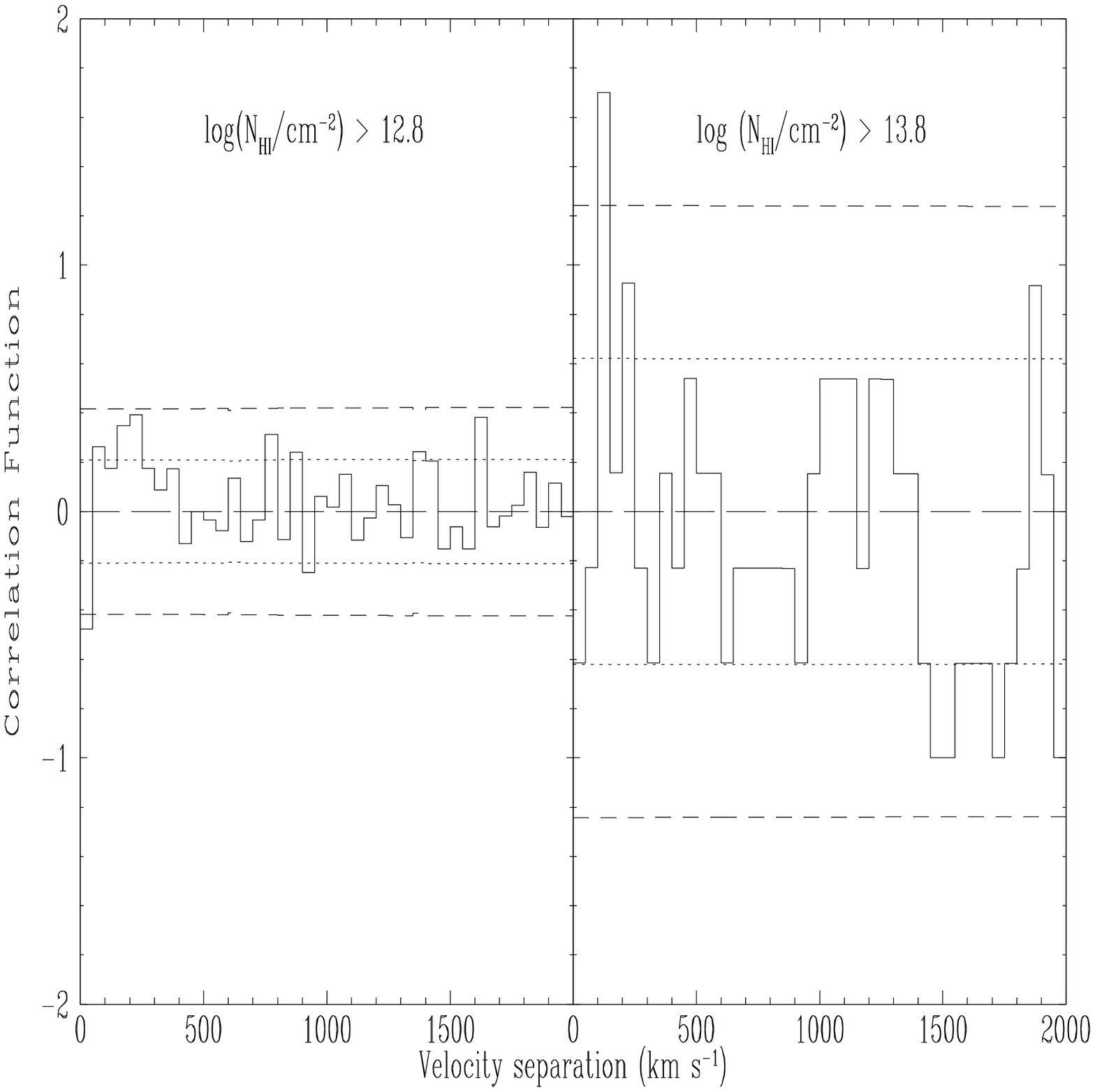}}
\end{picture}
\caption[]{\label{corr1} The correlation function for clouds obtained from 
simulations with 
$\gamma=1.7$, $T_m=3.01 \times 10^4$K, $f=0.026^2$ 
and $T_0=1.5 \times 10^4$K. 
The results are presented for two different column density thresholds. 
The dotted and the short-dashed lines show the 1$\sigma$ and 2$\sigma$ 
deviation from random distribution, respectively. The velocity bin width used 
is 50 km s$^{-1}$.
}
\end{figure*}

The correlation function for the clouds $\xi_{\rm cloud}(\Delta v)$ 
obtained from our simulations is shown
in Figure \ref{corr1}, for two different column density thresholds. We have 
used a velocity bin of 50 km s$^{-1}$. 
We have 
also marked the $1\sigma$ and $2\sigma$ significance levels in the 
figure, obtained using a Poisson distribution.
We can see that there is virtually no correlation above $2\sigma$ 
significance level when the column density threshold is low 
($N_{\rm HI}>12.8$ cm$^{-2}$). 
There is a clear positive correlation (2.78$\sigma$ significance) 
in the velocity bin around $\Delta v= 125$ km s$^{-1}$ for clouds with 
$N_{\rm HI} > 10^{13.8}$cm$^{-2}$. 
The dependence of clustering on the 
strength of the lines was noted by Cristiani et al. (1997) 
and Srianand (1997).
Kim et al. (1997) find a positive 
correlation of 2.8$\sigma$ significance at the velocity bin 
50--100 km s$^{-1}$ in the observed data.
Cristiani et al. (1997) too find a positive 
correlation of about 7$\sigma$ significance at $\Delta v=100$ km s$^{-1}$, in 
a wide redshift range $1.7 <z<3.1$ using a much larger number of samples. 

While concentrating on the parameter values $f=0.026^2$ and 
$T_0=1.5\times 10^4$K, 
we would like to see the effect of $\gamma$ on the statistics obtained from 
the Voigt profile decomposition. 
The comparison of the $b$-distribution for different 
values of $\gamma$ is given in Table \ref{b_chi}. We have performed a 
$\chi^2$ test, the results being shown in the same table. The mean $b$ 
increases with $\gamma$, which is due to the fact that the range of 
temperatures is higher for large $\gamma$.

\begin{table}
\caption{Comparison between simulated $b$ distribution and observations 
for $f=0.026^2$, $T_0=1.5 \times 10^4$K and different values of $\gamma$. The
observed mean value of $b$ is 36.35 km s$^{-1}$ at $z=2.31$.}
\begin{tabular}{|c|c|c|}
\hline
$\gamma$ & $\chi^2/\nu$ (likelihood) & Mean $b$ (km s$^{-1}$)\\
\hline
1.5 & $0.70$ (90.2\%) & $34.90$\\
1.7 & $0.61$ (96.3\%) & $35.19$\\
2.1 & $0.92$ (59.5\%) & $37.39$\\
\hline
\end{tabular}
\label{b_chi}
\end{table}

Next, we compare the simulated 
column density distribution, $f(N_{\rm HI})$ with observations for different 
values of $\gamma$. 
Table \ref{fnh} gives the values of the slope of the distribution,
$\beta_{HI}$ in the column density
range $12.8\le\log(N_{HI}/{\rm cm}^{-2})\le16.0$ for different $\gamma$. 
The distribution becomes 
steeper as we increase $\gamma$. 
This is consistent with semi analytic results of Paper I.
It is very clear that the slope for 
$\gamma<1.5$ ($\gamma>2.1$) is too flat (steep) 
to match the observations, even within error limits. This is somewhat 
consistent with what we find from the $b$-distribution above.
We have checked and found that the ${\rm d}N/{\rm d}z$ for 
$13.77\le\log(N_{HI}/{\rm cm}^{-2})\le16.0$ 
for different values 
of $\gamma$ is well within the observed range.

\begin{table}
\caption{Comparison between simulated column density distribution and 
observations. The
observed value of $\beta_{HI}$ is around 1.35 at $z=2.31$ in the column
density range $12.8\le\log(N_{HI}/{\rm cm}^{-2})\le16.0$ (Kim et al 1997). The 
value of $f$ and $T_0$ are $0.026^2$ and $1.5\times 10^4$K respectively.
}
\begin{tabular}{|c|c|}
\hline
$\gamma$ & $\beta_{HI}$\\
\hline
1.5 & $1.24\pm 0.12$\\
1.7 & $1.31\pm 0.13$\\
2.1 & $1.44\pm 0.14$\\
\hline
\end{tabular}
\label{fnh}
\end{table}

Finally, we discuss the correlation functions. Table \ref{corr_gamma} 
shows the correlation function within 
100 km s$^{-1} < \Delta v < 150$ km s$^{-1}$ 
for different values of $\gamma$. There is 
a slight increase in the correlation amplitude as one increases $\gamma$. 
One can, in principle, use this trend to constrain the value of $\gamma$ 
through correlation function. However, here we cannot do so because of the 
large errors ($\sigma \sim 0.6$). 
We have taken the same number of lines as is done
in the observations (about 100--140). Consequently, the errors are large and 
the correlation functions 
are consistent with observations for a wide range of parameter 
values. 
\begin{table}
\caption{The correlation function for 
Ly$\alpha$ clouds within 100 km s$^{-1} < \Delta v < 150$ km s$^{-1}$
for $f=0.026^2$, $T_0=1.5\times 10^4$K and different values of 
$\gamma$, obtained from the simulations. }
\begin{tabular}{|c|c|}
\hline
$\gamma$ & $\xi_{\rm cloud}(100$ km s$^{-1} < \Delta v < 150$ km s$^{-1}$)\\
\hline
1.5 & $1.42$\\
1.7 & $1.70$\\
2.1 & $2.65$\\
\hline
\end{tabular}
\label{corr_gamma}
\end{table}

To summarise, we have shown that, for some particular parameter range, 
our model is 
consistent with all of the observations (within error limits) obtained 
from the transmitted flux and from the Voigt profile decomposition 
of the observed spectrum. 
This justifies our approach of modelling the IGM using the 
lognormal approximation.
We have also shown that it is possible to 
put stringent limits on the $\gamma-T_0$ plane for a given $f$ using 
transmitted flux statistics only.

\section{Discussions and Summary}

We have performed a simulation of the Ly$\alpha$
absorption spectrum originating from the low density IGM using a 
semi analytic ansatz. We have
studied the effect of various parameters on the spectrum and the
concerned statistics. We have found that our
simulations match most of the observations available for a 
narrow parameter range.

(i) Various statistics performed on the
simulated data  and the observed points provided by McDonald (2000a)
over a redshift range 2.09$-$2.67, constrain the value of $f$ within 
$0.020^2$--$0.032^2$, independent of $T_0$ and $\gamma$. In this range of 
$f$, we considered three particular values of $f$, namely, $0.023^2$, 
$0.026^2$ and $0.029^2$. We 
constrain $T_0$ within (0.8--2.5)$\times 10^4$K and $\gamma$ within 
1.3--2.3. If the 
range in $f$ is narrowed down through some other studies, the values of 
$\gamma$ and $T_0$ can be constrained further. 
Although the observations allow $f$ in the range $0.020^2$ to $0.032^2$, 
we find that the match between simulations and observations is best for 
$f\sim 0.026^2$.

\begin{figure}
\psfig{figure=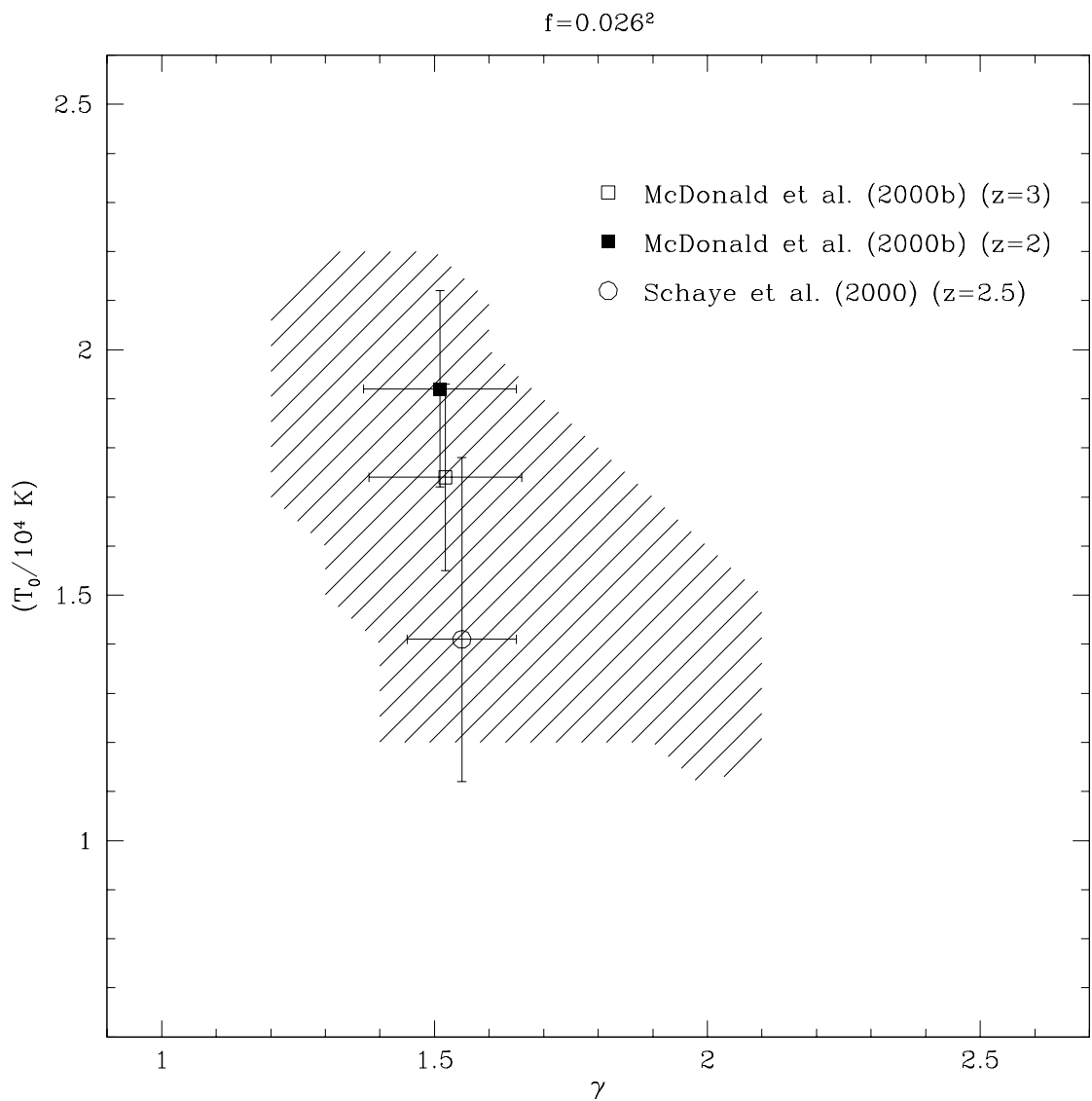,width=16cm}
\caption{\label{t0_gamma_comp}
Comparison of the bounds on $\gamma$ and $T_0$ obtained by us with other
results. The shaded region shows the bound from this work for $f=0.026^2$. 
The open and the filled squares denote the parameter values obtained by 
McDonald et al. (2000b) depending on whether they use 
the output from hydrodynamical simulations 
at $z=3$ or $z=2$, respectively. The value obtained 
by Schaye et al. (2000) for QSO Q1442 around $z=2.5$ is denoted by the open 
circle.
}
\end{figure}

The values of $T_0$ and $\gamma$ are usually obtained (in previous attempts) 
from observational 
data through the Voigt profile fitting and the lower envelope 
$N_{\rm HI}-b$ scatter plot. 
The range obtained by us is consistent with the one obtained by 
Schaye et al. (2000).   
They infer $1.26\times 10^4$K$<T_0< 2.00\times 
10^4$K and $\gamma=1.45-1.65$ for  the spectrum of 
QSO Q1442 at $z=2.5$ (see their Figure 6). 
McDonald et al. (2000b) 
use the lower cutoff of the $N_{\rm HI}-b$ scatter plot  
to infer $T_0$ and $\gamma$.
For $\bar{z}=2.4$, they find 
$T_0=(1.74 \pm 0.19) \times 10^4$K,$\gamma=1.52\pm0.14$ or 
$T_0=(1.92 \pm 0.2) \times 10^4$K,$\gamma=1.51\pm0.14$, 
depending on whether they calibrate the data using 
the output from hydrodynamical simulations 
at $z=3$ or $z=2$, respectively. 
On the other hand, the hydrodynamical simulations (McDonald et al. 2000a) 
give slightly lower values of $T_0$, i.e., 
$T_0=1.31\times 10^4$K and $1.6\times 10^4$K, for $z=2$ and $z=3$, 
respectively. 
Ricotti et al. (2000) 
also use the $N_{\rm HI}-b$ scatter plot to 
constrain $T_0$ between (1--2.4)$\times 10^4$K 
at $z=1.90$ and between (2--2.7)$\times 10^4$K at $z=2.75$. The corresponding 
constraints on $\gamma$ are $1.32\pm0.30$ at $z=1.90$ and $1.22\pm 0.10$ 
at $z=2.75$. 
For clarity, 
we show how our results compare with those obtained by other people in 
Figure \ref{t0_gamma_comp}. 
It should be clear from the figure and from 
the above discussion that our results are in quite good agreement 
with others.

It should, however, be stressed that 
the results obtained using the Voigt profile decomposition and 
$N_{\rm HI}-b$ scatter plot have certain inherent biases when 
compared with those obtained from the transmitted flux. 
For low column density clouds, the error is introduced in 
the $b$ values because of the noise in the spectrum. Also considerable 
fraction of low column density low $b$ lines 
that may not trace the low density regions are artificially introduced to 
get a better $\chi^2$ while fitting the blends of saturated lines. For the
high column density clouds, due to saturation, there is a degeneracy 
between velocity dispersion and number of components to be fitted. Our 
constraints on $\gamma$ and $T_0$ using the transmitted flux statistics are 
free from the above mentioned effects.
These constraints on $\gamma$ and $T_0$ can be used 
simultaneously to constrain the reionisation epoch and the reionisation 
temperature (Hui \& Gnedin 1997). 

(ii) We constrain $f=(\Omega_B h^2)^2/J_{-12}$ to be in the 
range $0.020^2$--$0.032^2$ regardless of the values of $T_0$ and $\gamma$. 
The values for $f$ found by McDonald et al. (2000a) 
are $(0.0257 \pm 0.0017)^2$ 
(for $T_0=1.31\times 10^4$K) and 
$(0.0239 \pm 0.0016)^2$ (for $T_0=1.6\times 10^4$K) for $z=2$ and $z=3$, 
respectively. 
This is consistent with the range found in our study. 

\begin{figure}
\psfig{figure=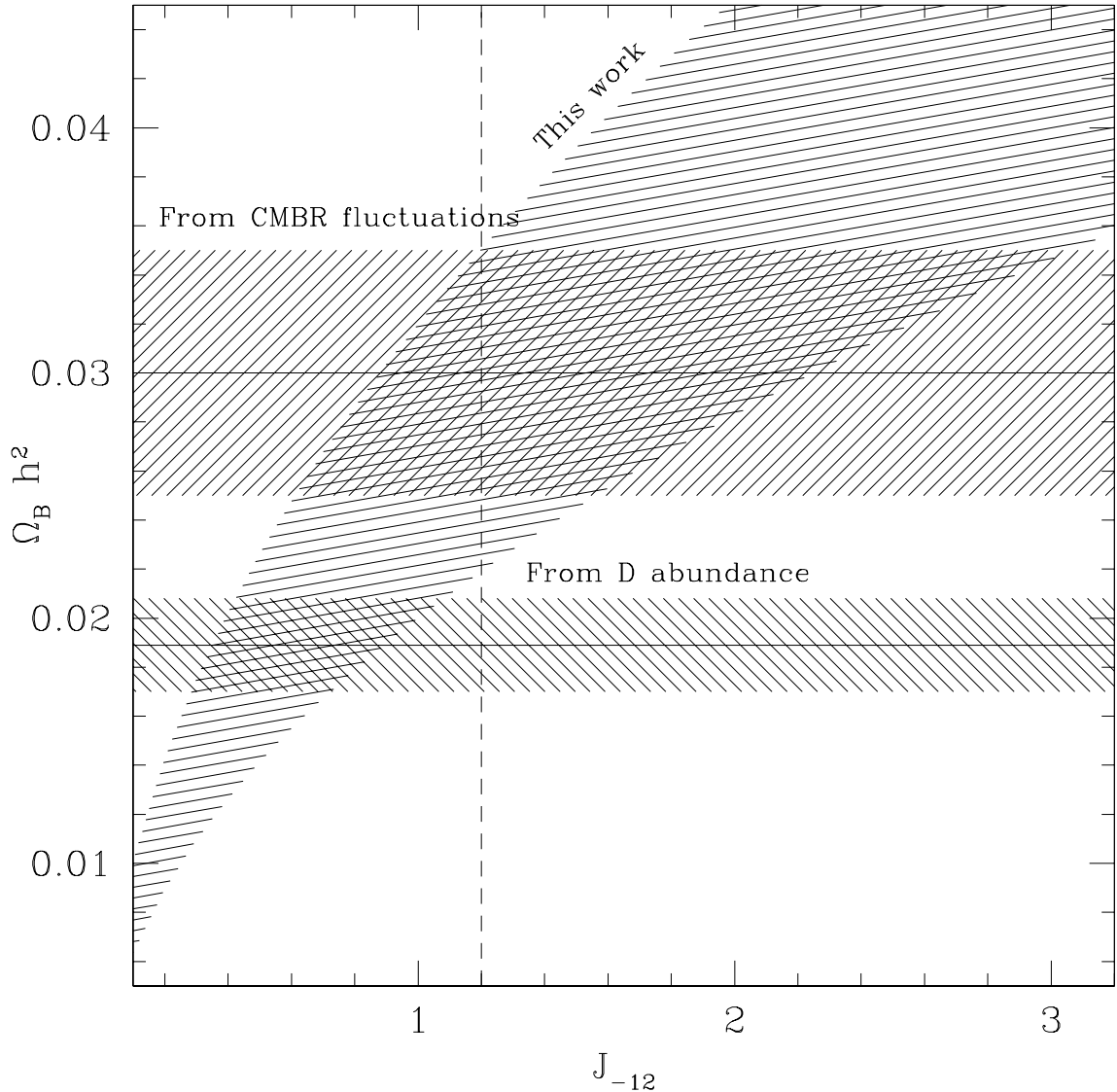,width=16cm}
\caption{\label{bound}
Comparison of the bounds on $\Omega_B h^2$ obtained from our simulations with 
those obtained from Big Bang Nucleosynthesis (BBN) and CMBR analyses. 
The lower 
horizontal bound shows the region allowed by BBN 
(Burles, Nollett \& Turner 2000) while the upper band shows that allowed 
by initial 
BOOMERANG and MAXIMA data (Bond et al. 2000; Padmanabhan \& Sethi 2000). 
The bound on $\Omega_B h^2$ arising from the current work is shown
as a function of $J_{-12}$ by the curved band running from left bottom to  
the right top. 
}
\end{figure}

The constraint we have obtained on $f$ is important because of the bound  
it implies on the baryon fraction of the universe. The situation is
illustrated in Figure \ref{bound}, 
where we plot $\Omega_B h^2$ as a function 
of $J_{-12}$.
The lower horizontal band corresponds to $0.0170<\Omega_B h^2 <0.0208$, 
which is considered to be the acceptable range of values from 
Big Bang Nucleosynthesis (BBN) (Burles, Nollett \& Turner 2000). As has
been noted by several authors and emphasised by Padmanabhan \& Sethi (2000), 
this is
already in contradiction with the 95 per cent 
confidence limits on $\Omega_B h^2$    
arising from the analysis of the initial BOOMERANG and MAXIMA data 
The latter bound ($0.025<\Omega_B h^2 <0.035$) is shown 
in the upper horizontal band in Figure \ref{bound} 
(for details see Bond et al. 2000; Padmanabhan \& Sethi 2000). 
The bound on $\Omega_B h^2$ arising from the current work 
$(0.020<\Omega_B h^2/\sqrt{J_{-12}}<0.032)$ is shown
as a function of $J_{-12}$ by the curved band running from left bottom to  
the right top. 
It is clear from the figure that if $J_{-12}>1.2$ 
(indicated by the vertical dashed line in the figure),
we have 
$\Omega_B h^2 >0.022$, which is in violation of BBN value. Haardt \& 
Madau (1996), using the QSO luminosity function, 
have estimated $J_{-12}=1.63$ for 
$\Omega_m=0.2$ open universe and $J_{-12}=1.13$ for 
$\Omega_m=1.0$ flat universe at $z=2.41$. This can be considered as a strict 
lower bound on $J_{-12}$, 
as galaxies also contribute equally to the ionising UV background at these 
redshifts (Steidel, Pettini \& Adelberger 2000). 
The bounds obtained from the proximity effect are $0.9<J_{-12}<3.1$ 
(Scott et al. 2000).
Thus, it appears that 
the bounds on $\Omega_B h^2$ obtained from the 
Ly$\alpha$ forest analysis could possibly be 
inconsistent with those obtained from the BBN.

While this paper was being
refereed, three groups have released further data 
(Netterfield et al. 2001,Pryke et al. 2001 and  Stompor et al. 2001)
 with some
inital analysis of their implications. The BOOMERANG group has given the
best bet values of $\Omega_B h^2\approx 0.02$ which is consistent with
BBN results (Netterfield et al. 2001). However, other group
still obtains $\Omega_B h^2\approx 0.03$ (Stompor et al. 2001).
There has also been a suggestion (Pettini\& Bowen, 2001) that the
a reanalysis of deutrium abundance might raise the BBN bound upwards to
about $\Omega_B h^2\approx 0.025$. The situation is therefore unclear at
present and these aspects must be kept in mind while assessing the
importance of the results in Fig.~\ref{bound}.
As far as the studies on IGM are concerned, we believe that 
it is important to estimate the value 
of $J_{-12}$ more rigorously so as to put a strong constraint on 
$\Omega_B h^2$. Further work in this direction is in progress.

\section*{Acknowledgment}

We gratefully acknowledge the support from the Indo-French Centre for 
Promotion of Advanced  Research under contract No. 1710-1.
TRC acknowledges financial support from 
the University Grants Commission, India. We also thank McDonald et al. (2000) 
for making the observational data public.


\begin{thebibliography}{99}

\bibitem[]{berg}
Bergeron J., Boisse P., 1991, A\&A, 243, 344
\bibitem[]{bi93} 
Bi H., 1993, ApJ, 405, 479
\bibitem[]{bd}
Bi H., Davidsen A. F., 1997, ApJ, 479, 523 (BD)
\bibitem[]{bbc}
Bi H. G., B\"{o}rner G., Chu Y., 1992, A\&A, 266, 1
\bibitem[]{bgf}
Bi H., Ge J., Fang L., 1995, ApJ, 452, 90
\bibitem[]{black}
Black J. H., 1981, MNRAS, 197, 553
\bibitem[]{bond1}
Bond J. R., Szalay A. S., Silk J., 1988, ApJ, 324, 627
\bibitem[]{bond2}
Bond J. R., Ade P., Balbi A., Bock J., Borrill J., Boscaleri A., Coble K., 
Crill B., et al., 2000, preprint (astro-ph/0011378)
\bibitem[]{bryan}
Bryan G. L., Machacek M. E., 2000, ApJ, 534, 57
\bibitem[]{burles}
Burles S., Nollett K. M., Turner M. S., 2000, preprint (astro-ph/0010171)
\bibitem[]{carlberg}
Carlberg R. G., Couchman H. M. P., 1989, ApJ, 340, 47
\bibitem[]{cen}
Cen R. Y., Miralda-Escud\'e J., Ostriker J. P., Rauch M. R., 1994,
ApJ, 437, L9
\bibitem[]{rps}
Choudhury T. R., Padmanabhan T., Srianand R., 2000, MNRAS, in press 
(preprint astro-ph/0005252) (Paper I)
\bibitem[]{coles1}
Coles P., Jones B., 1991, MNRAS, 248, 1
\bibitem[]{coles2}
Coles P., Melott A. L., Shandarin S. F., 1993, MNRAS, 260, 765
\bibitem[]{croft98}
Croft R. A. C., Weinberg D. H., Katz N., Hernquist L., 1998, ApJ, 495, 44
\bibitem[]{croft99}
Croft R. A. C., Weinberg D. H., Pettini M., Hernquist L., Katz N.,
1999, ApJ, 520, 1
\bibitem[]{cris}
Cristiani S., D'Odorico S., D'Odorico V., Fontana A., Giallongo E., Moscardini
L., Savaglio S., 1997, in Petitjean P., Charlot S. ed.,
Proc. of the 13$^{th}$ IAP Astrophysics Colloquium,
Structure and Evolution of the Intergalactic 
Medium from QSO Absorption Line Systems. Editions Fronti\`{e}res,
Paris, p. 165
\bibitem[]{dave} 
Dav\'e R., Hernquist L., Katz N., Weinberg D. H., 1999, ApJ, 511, 521
\bibitem[]{dorosh}
Doroshkevich A. G., Shandarin S. F., 1977, MNRAS, 179, 95
\bibitem[]{ebw} 
Efstathiou G., Bond J. R., White S. D. M., 1992, MNRAS, 258, 1P
\bibitem[]{eke}
Eke V. R., Cole S., Frenk C. S., 1996, MNRAS, 282, 263
\bibitem[]{fang}
Fang L. Z., Bi H., Xiang S., B\"{o}rner G., 1993, ApJ, 413, 477
\bibitem[]{gnedin}
Gnedin N. Y., 1998, MNRAS, 299, 392
\bibitem[]{gh}
Gnedin N. Y., Hui L., 1996, ApJ, 472, L73
\bibitem[]{hm}
Haardt F., Madau P., 1996, ApJ, 461, 20
\bibitem[]{hern}
Hernquist L., Katz N., Weinberg D. H., Miralda-Escud\'e J., 1996, ApJ, 457,
L51
\bibitem[]{hui}
Hui L., 1999, ApJ, 516, 519
\bibitem[]{hg} 
Hui L., Gnedin Y. G., 1997, MNRAS, 292, 27
\bibitem[]{hgz}
Hui L., Gnedin Y. G., Zhang Y., 1997, ApJ, 486, 599
\bibitem[]{khare} 
Khare P., Srianand R., York D. G., Green R., Welty D.,
Huang K., Bechtold J., 1997, MNRAS, 285, 167
\bibitem[]{kim}
Kim T., Hu E. M., Cowie L. L., Songaila A., 1997, AJ, 114, 1
\bibitem[]{kofman}
Kofman L., Gnedin N., Bahcall N., 1993, ApJ, 413, 1
\bibitem[]{kulkarni}
Kulkarni V. P., Huang K., Green R. F., Bechtold J., Welty D. E., 
York D. G., 1996, MNRAS, 279, 197
\bibitem[]{lomb}
Lomb N. R., 1976, Ap\&SS, 39, 447
\bibitem[]{mcdonald1}
McDonald P., Miralda-Escud\'e J., Rauch M., Sargent W. L. W.,
Barlow T. A., Cen R., Ostriker J. P., 2000a, ApJ, 543, 1
\bibitem[]{mcdonald2}
McDonald P., Miralda-Escud\'e J., Rauch M., Sargent W. L. W.,
Barlow T. A., Cen R., 2000b, preprint (astro-ph/0005553)
\bibitem[]{mcgill}
Mcgill C., 1990, MNRAS, 242, 544
\bibitem[]{miralda}
Miralda-Escud\'e J., Cen R., Ostriker J. P., Rauch M., 1996, ApJ, 471, 582
\bibitem[]{boom}
Netterfield C. B., Ade P. A. R., Bock J. J., Bond J. R., et al., 2001, 
preprint (astro-ph/0104460)
\bibitem[]{nitya}
Nityananda R., Padmanabhan T., 1994, MNRAS, 271, 976
\bibitem[]{ostri}
Ostriker J. P., Steinhardt P. J., 1995, Nat, 377, 600
\bibitem[]{paddy}
Padmanabhan T., 1996, MNRAS, 278, L29
\bibitem[]{padsethi}
Padmanabhan T., Sethi S. K., 2000, Ap. J (in press); (astro-ph/0010309)
\bibitem[]{pettini}
Pettini, M., Bowen, D. V. 2001, preprint (astro-ph/0104474)
\bibitem[]{press}
Press W. H., Teukolsky S. A., Vetterling W. T., Flannery B. P., 1992, 
Numerical recipes in FORTRAN. 
Cambridge: University Press
\bibitem[]{dasi}
Pryke C., Halverson N. W., Leitch E. M., Kovac J., Carlstrom J. E., 2001, 
preprint (astro-ph/0104490)
\bibitem[]{rauch}
Rauch M., Miralda-Escud\'e J., Sargent W. L. W., Barlow T. A.,
Weinberg D. H., Hernquist L., Katz N., Cen R., et al., 1997, ApJ, 489, 7
\bibitem[]{ricotti}
Ricotti M., Gnedin N. Y., Shull J. M., 2000, ApJ, 534, 41
\bibitem[]{rie}
Riediger R., Petitjean P., M\"ucket J. P., 1998, A\&A, 329, 30
\bibitem[]{sargent}
Sargent W. L. W., Young P. J., Boksenberg A., Tytler D., 1980, ApJS, 42, 41
\bibitem[]{scargle}
Scargle J. D., 1982, ApJ, 263, 835
\bibitem[]{schaye_99}
Schaye J., Theuns T., Leonard A., Efstathiou G., 1999, MNRAS, 310, 57
\bibitem[]{schaye_00}
Schaye J., Theuns T., Rauch M., Efstathiou G., Sargent W. L. W., 2000,
MNRAS, 318, 817
\bibitem[]{scott}
Scott J., Bechtold J., Dobrzycki A., Kulkarni V. P., 2000, ApJS, 130, 67
\bibitem[]{anand1} 
Srianand R., 1996, ApJ, 462, 68
\bibitem[]{anand2} 
Srianand R., 1997, ApJ, 478, 511
\bibitem[]{ak1} 
Srianand R., Khare P., 1994, MNRAS, 271, 81
\bibitem[]{ak2} 
Srianand R., Khare P., 1996, MNRAS, 280, 767
\bibitem[]{stei}
Steidel C. C., 1993, in Shull J. M., Thronson H. A., eds, Proc. of the 3rd
Teton Astronomy Conference, The Environment and Evolution of Galaxies.
Dordrecht, Kluwer, p. 263
\bibitem[]{stei2}
Steidel C. C., Pettini M., Adelberger K. L., 2000, 
preprint (astro-ph/0008283)
\bibitem[]{maxima}
Stompor R., Abroe M., Ade P., Balbi A., et al., 2001, 
preprint (astro-ph/0105062)
\bibitem[]{tle}
Theuns T., Leonard A., Efstathiou G., 1998, MNRAS, 297, L49
\bibitem[]{theuns}
Theuns T., Leonard A., Efstathiou G., Pearce F. R., Thomas P. A.,
1998, MNRAS, 301, 478
\bibitem[]{webb}
Webb J. K., 1987, PhD thesis, Cambridge Univ.
\bibitem[]{zhang95}
Zhang Y., Anninos P., Norman M. L., 1995, ApJ, 453, L57
\end{thebibliography}
\end{document}